\documentclass[aps,showpacs,twocolumn,superscriptaddress,nofootinbib]{revtex4-1} \input epsf
\usepackage{graphicx}
\usepackage{color}
\usepackage{dcolumn}
\usepackage{bm}
\usepackage{amsmath}
\usepackage[normalem]{ulem}
\usepackage{multirow}

\begin{document}
\preprint{APS/123-QED}
\title{A lattice model of ternary mixtures of lipids and cholesterol with tunable domain sizes}
\author{Tanmoy Sarkar}
\affiliation{Department of Biomedical Engineering, Ben Gurion University of the Negev, Be'er Sheva 84105, Israel}
\author{Oded Farago}
\email{ofarago@bgu.ac.il}
\affiliation{Department of Biomedical Engineering, Ben Gurion University of the Negev, Be'er Sheva 84105, Israel}
\date{\today}
\begin{abstract}

{Much of our understanding of the physical properties of raft domains
  in biological membranes, and some insight into the mechanisms
  underlying their formation stem from atomistic simulations of simple
  model systems, especially ternary mixtures consisting of saturated
  and unsaturated lipids, and cholesterol.  To explore the properties
  of such systems at large spatial scales, we here present a simple
  ternary mixture lattice model, involving a small number of nearest
  neighbor interaction terms. Monte Carlo simulations of mixtures with
  different compositions show an excellent agreement with experimental
  and atomistic simulation observations across multiple scale, ranging
  from the local distributions of lipids to the phase diagram of the
  system. The simplicity of the model allows us to identify the roles
  played by the different interactions between components, and the
  interplay between them. Importantly, by changing the value of one of
  the model parameters, we can tune the size of the liquid-ordered
  domains, thereby to simulate both Type II mixtures exhibiting
  macroscopic phase separation and Type I mixtures with nanoscopic
  domains. The Type II mixture simulation results fit well to the
  experimentally-determined phase diagram of mixtures containing
  saturated DPPC/unsaturated DOPC/Chol. When the tunable parameter is
  changed, we obtain the Type I version of DPPC/DOPC/Chol, i.e., a
  mixture not showing thermodynamic phase transitions but one that may
  be fitted to the same phase diagram if local measures are used to
  distinguish between the different states. Our model results suggest
  that short range packing is likely to be a key regulator of the
  stability and size distribution of biological rafts.}

\end{abstract}


\maketitle

\section*{Introduction}
\label{sec:intro}

Biological membranes are complex mixtures of many types of lipids,
proteins and cholesterol (Chol)~\cite{{alberts-book}}. Though the
lateral organization of such membranes is far from being fully
elucidated, evidences indicate that it is highly
heterogeneous~\cite{levental20}. Specifically, biological membranes
contain small (10-200 nm) and dynamic domains, known as rafts, that
are enriched in saturated lipids, cholesterol (Chol) and often
particular proteins~\cite{simons97,pike06,levental20}.  Rafts are
liquid-ordered ($L_o$) domain ``floating'' in a sea of
liquid-disordered ($L_d$) lipids~\cite{vdgoot01,kaiser09}. The $L_d$
and $L_o$ phase are liquid crystalline ones, where the lipids diffuse
in the membrane plane~\cite{filippov04}. In the $L_o$ phase, the
hydrocarbon chains are ordered, fully extended and tightly packed,
similarly to the gel ($S_o$) phase~\cite{holl08}. Because of the
enormous complexity of biological membranes, much of our understanding
of the biophysical behavior and properties of raft domains is based on
investigations of simple model membranes with few components,
especially ternary mixtures of saturated and unsaturated lipids and
Chol~\cite{feigenson09,goni08,veatch05,levental16,chong}.  Studies of
many different ternary mixtures reveal a phase diagram with regions of
coexisting phases in the composition space, including coexistence
between $L_o$ and $L_d$ phases which is believed to have relevance to
raft domains in biological membranes~\cite{feigenson09, komura14,
  veatch07, hirst11}. Depending on the identity of the lipids and
temperature, the $L_d$ and $L_o$ regions in ternary mixtures may be
arranged in one of two ways: In mixtures like
DPPC/DOPC/Chol~\cite{veatch03} or DSPC/DOPC/Chol~\cite{zhao07}, for
instance, they are thermodynamically (macroscopically) phase separated
over a wide range of compositions. In contrast, no true phase
separation is observed when the doubly-unsaturated DOPC in these
ternary mixtures is replaced with POPC, which has only one unsaturated
hydrophobic chain.  Instead, the liquid-ordered regions form small
domains that are surrounded by a liquid-disordered matrix. This
arrangement is commonly referred to as {\em microscopic phase
  separation}\/ - a term that we also adopt here, emphasizing that it
does not describe coexistence of phases in the thermodynamic
sense. Feigenson~\cite{feigenson09} termed mixtures exhibiting
thermodynamic (macroscopic) and ``microscopic'' phase separation as
Type II and Type I mixtures, respectively. Type II mixtures show
visible domains in fluorescence microscopy which makes their
identification easy.  In contrast, Type I mixtures appear uniform, and
the presence of liquid-ordered domains in these systems must be
inferred by indirect spectroscopic methods.

Generally speaking, formation of small and transient domains may be
associated with a vanishingly small line tension between the liquid
phases. In Type II mixtures, this is expected in the one phase region
of the phase diagram close to the critical demixing
point~\cite{veatch07}. In the two phase region, a comparatively higher
line tension drives the macroscopically large segregation of the
liquid-ordered phases.  Additionally, there are also other mechanisms
that may lead to the appearance of small domains, including in Type I
mixtures not exhibiting macroscopic phase separation (see recent
reviews in~\cite{komura14,schmid17review}), e.g., two dimensional
microemulsion domains stabilized by line active molecules (lineactant)
like hybrid lipids~\cite{brewster09,palmieri14},
or a coupling between the local lipid composition and
bilayer~\cite{leibler,schick1,sadeghi} or monolayer~\cite{meinhardt13}
curvatures. All of these mechanisms may be relevant to the formation
of raft domains in biological membranes that are compositionally more
complex and typically larger than liquid-ordered domains in Type I
mixtures~\cite{schmid17review}.

Accessing the length and time scales relevant to lipid domains in
complex membranes is computationally very challenging. Atomistic
simulation is the putative method for characterizing the molecular
organization in lipid membranes~\cite{enkavi}. It provides a molecular
scale picture which is not available experimentally. However, even for
simple model membranes, all-atom large scale simulations are
computationally too costly. Therefore, a coarse-grained (CG)
description is employed to allow access to the larger scales. In the
past 15 years, CG simulations involving the MARTINI force-field have
become very popular in the membrane biophysics
community~\cite{marrink}.  In MARTINI simulations, several atoms are
unified and represented by effective beads.  One particular problem of
MARTINI simulations which often leads to inconsistencies with
atomistic simulations of similar system, is the chain entropy which is
heavily involved in the phase transition but is not properly
represented when using a smaller number of effective beads (see review
in~\cite{enkavi} for examples of discrepancies between atomistic and
MARTINI simulations of lipid-Chol mixtures).  Beyond CG MARTINI (and
similar force fields) simulations of {\em specific}\/ lipids, we have
more (ultra) CG {\em generic} models designed for studying the general
mechanisms and biophysical driving forces governing the thermodynamic
behavior of lipid mixtures~\cite{meinhardt19}. The highest degree of
abstraction and computational efficiency is offered by lattice
models~\cite{ispen87,pink80, almeida11, joannis,heberle,
  dai,reigada,banerjee,eckstein, frazier}. Lattice models have been
extensively used as a tool for studying phase transitions, and many
predictions of these models have proven to be relevant to lipid
mixtures~\cite{keller09,almeida11}. Recently, we presented a lattice
model for a binary mixture of the saturated lipid DPPC and Chol, that
correctly captures the thermodynamic behavior of the system across
multiple scales~\cite{tanmoy}. On the macroscopic scales, the lattice
simulations reproduced the well-established phase diagram of DPPC/Chol
mixtures, including regions where liquid-ordered domains of size
$\sim$ 10-100 nm are observed in a liquid-disordered matrix. On the
molecular scales, the simulations reveal the existence of gel-like
nano clusters of size $\sim$ 1-10 nm within the larger Chol-rich
liquid-ordered domains. Similar sub-structures have been seen in
atomistic simulations of this system~\cite{javanainen17}.

\begin{figure*}
\centering
\includegraphics[width=16cm]{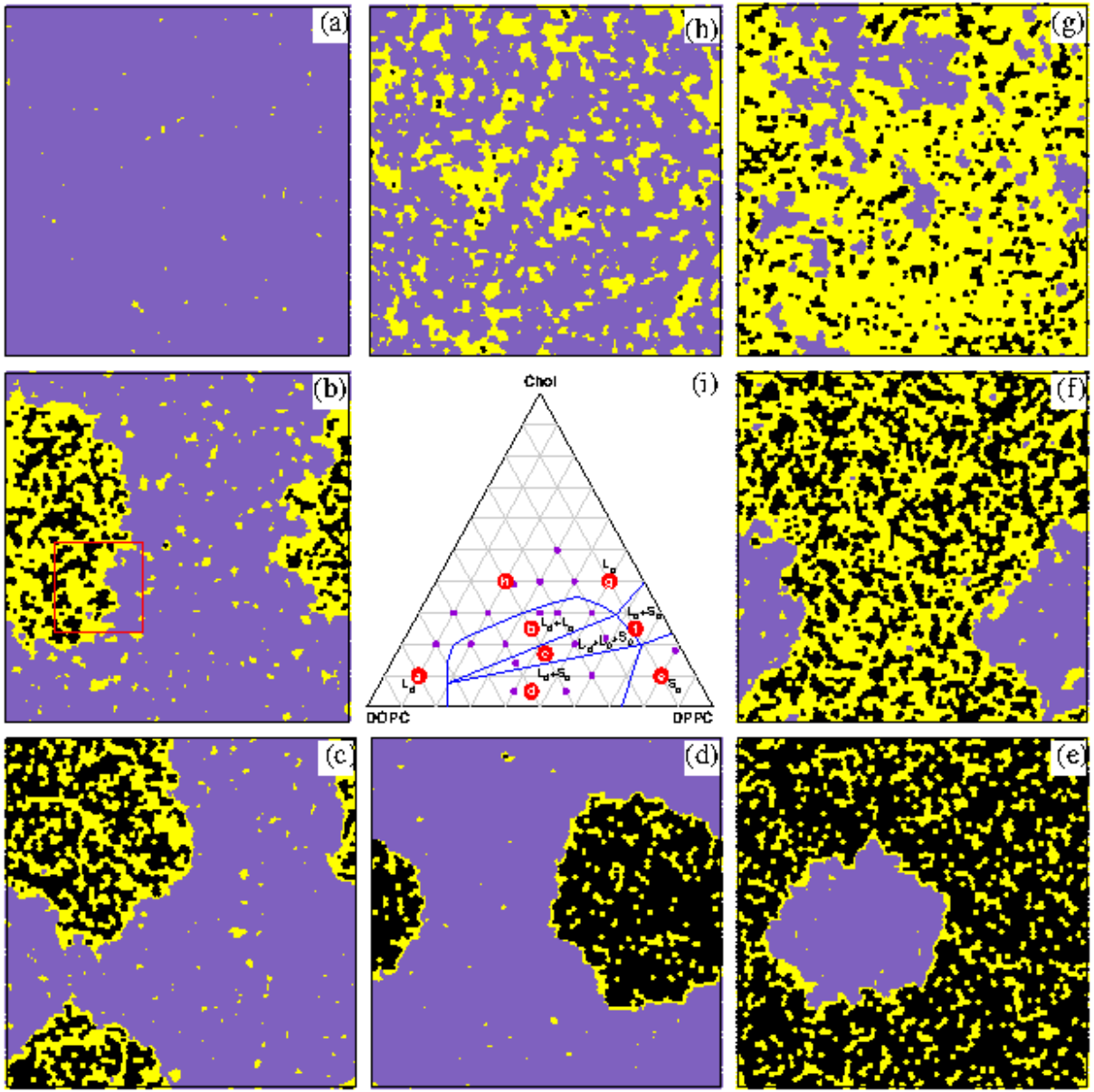}
\caption{Phase diagram and snapshots of the lateral organization of
  DPPC/DOPC/Chol mixtures at different compositions. The
  liquid-disordered ($L_d$), liquid-ordered ($L_o$), and gel ($S_o$)
  sites are colored in purple, yellow, and black, respectively. The
  snapshots are taken at temperature $T=280$ K and for
  $\epsilon_{24}=0$ . Figure (a), (b), (c), (d), (e), (f), (g) and (h)
  show, respectively, snapshots from $L_d$, $L_d+L_o$, $L_d+L_o+S_o$,
  $L_d+S_o$, $S_o$, $L_o+S_o$, $L_o$ and a single heterogeneous phase
  containing liquid-ordered domains in a liquid-disordered
  membrane. (i) Phase diagram of the ternary mixtures at $T=283K$. The
  blue phase separating lines are adapted from
  ref.~\cite{veatch07}. The purple colored dots show the compositions
  of the simulated systems. The red dots with letters indicate the
  compositions of the snapshots.}
\label{fig1}
\end{figure*}

Fig.~\ref{fig1}(i) shows the phase diagram of DPPC (saturated)/DOPC
(doubly unsaturated)/Chol ternary mixture at $T=283$ K. This canonical
model system for studying the formation of liquid-ordered
domains~\cite{veatch07,veatch03,davis,uppamoochikkal} has been
identified as a Type II mixture. Depending on the mole fractions of
the lipids and Chol, the mixture may be found in a single or
coexisting phases.  Other Type II mixtures also show similar phase
diagrams, where the exact location of the phase separating lines
depends on the identity of the lipids and the temperatures.
Curiously, many Type I ternary mixtures also feature phase diagrams
resembling Fig.~\ref{fig1}~(i)~\cite{heberle10,dealmeida}, where the
liquid phases are microscopically rather than macroscopically
separated. The presence of small ordered domains in Type I mixtures is
inferred from studies based on different experimental methods, e.g.,
NMR~\cite{veatch07}, small-angle neutron scattering~\cite{Heberle13},
X-Ray diffraction~\cite{clarke}, and fluorescence resonance energy
transfer (FRET)~\cite{frazier}. These observations can be
intinterpeted in one of two ways: (i) a coexistence between two phases
with extremely small line tension, or (ii) a single phase with local
heterogeneities. From a theoretical point of view, the difference
between these two interpretations is marginal, but if the latter
interpretation is adopted then the liquid-liquid coexistence region in
the phase diagram does not really exist. This controversy
provides some of the motivation for the development of the model,
presented herein, of ternary mixtures with tunable domain sizes. In
the following, we present a minimal lattice model for a Type II
mixture that fits very well the DPPC/DOPC/Chol phase diagram. The
model contains a very small number of parameters, one of which
directly controls the line tension between the disordered and ordered
lipids. We demonstrate that by tuning the value of this parameter, we
can change the nature of the mixture from Type II to Type I, i.e.,
convert macroscopically separated liquid phases into a mixture with
traces of inhomogeneity at the local scale.

In atomistic and CG simulations, the presence of liquid-ordered
domains is often based on visual inspection of the lateral
organization and chain conformations of the lipids and Chol. One
particular problem in computer simulations is the fact that the
liquid-ordered domains may themselves be heterogeneous and contain
nanoscopic gel-like clusters. The presence of such Chol-free
hexagonally-packed clusters of acyl chains in liquid-ordered regions
that are rich in Chol have been observed in
simulations~\cite{javanainen17, tanmoy} of binary DPPC/Chol mixtures,
as well as in simulations of ternary DPPC/DOPC/Chol
mixtures~\cite{sodt}. This local heterogeneous distribution of
``domains within domains'' is also consistent with experimental
scattering data~\cite{amstrong13}. Another problem of detailed
computer simulations are finite size effects. A simulation study of a
ternary DPPC/DIPC/Chol mixture has pointed that the nature of the two
phase region and domain sizes may be strongly influenced by the size
of the simulated system~\cite{pantelopulos}. This may explain result
of recent atomistic simulations of DPPC/DOPC/Chol mixtures where phase
coexistence between liquid crystalline $L_d$ and $L_o$ is studied in a
system consisting of $\sim~1000$ lipids~\cite{tieleman20}. That study,
however, identified the system as a Type I mixture with domains of
characteristic size of the order of 10 nm, which is in disagreement
with clear experimental evidences indicating that this mixture is of
Type II~\cite{veatch03} featuring macroscopic phase separation.  This,
again, reinforces the importance of developing a computationally
simple lattice model that allows simulations at large spatial and
temporal scales. The attractiveness of lattice simulations lies not
only in their computational simplicity, but also in their minimal
nature which puts the focus on the factors that are essential for
understanding the biophysics of lipid mixtures and the mechanisms
governing their thermodynamic phase behavior.

The DPPC/DOPC/Chol lattice model presented below is an extension of
our recent successful lattice model of binary DPPC/Chol
mixtures~\cite{tanmoy}.  As will be shown below, the model provides a
multiscale picture of the ternary mixture over an unprecedented wide
range of scales. At the macroscopic scales, the model yields very nice
agreement with the experimentally-derived phase diagram of this
mixture, featuring all single and coexisting phases. At the
microscopic scales, the model unravels the inhomogeneous nano scale
distribution of lipids and Chol. Specifically, the model agrees with
atomistic simulations showing that the liquid-ordered domains contain
small gel cores of highly aligned DPPC chains~\cite{sodt}. The
simplicity of the model helps identifying that the origin of these
gel-like clusters can be attributed to the effective packing
interactions of the saturated chains with each other and with the
Chol. We further demonstrate that the strength of the packing
interaction between ordered DPPC and DOPC chains controls the size of
the domains. Thus, we can artificially eliminate the thermodynamic
$L_d+L_o$ phase coexistence in DPPC/DOPC/Chol mixtures in favor of
formation of smaller liquid-ordered domains. The resulting
locally-inhomogeneous Type I mixture may be viewed as either a single
phase or as thermally accessible configurations of a mixture of two
coexisting liquid phases with a negligible line tension between them,
and which quite surprisingly also fit the DPPC/DOPC/Chol phase
diagram.

\section*{Methods}
\label{sec:model}

The lattice model of ternary mixtures presented herein is based on the
model of DPPC/Chol binary mixture introduced in
ref.~\cite{tanmoy}. Each chain is represented as a lattice
point. Thus, the lipids with two acyl chains are represented as dimers
while the Chol molecules occupy a single site. All Monte Carlo (MC)
simulations are conducted on a triangular lattice of $N_s=121\times
140=16940$ (having an aspect ratio which is very close to unity) with
periodic boundary conditions. The simulations are performed on a
single planar lattice. This means that we do not explicitly introduce
mechanisms speculated to be relevant for raft formation such as
cross-monolayer interactions or curvature effects. A few lattice sites
are left empty, thereby allowing the molecules to diffuse on the
lattice which mimics the fluidity of the liquid membrane. The voids
also allow the variations in the local density of the chains, which
are expected in cases of coexistence between different phases. As in
ref.~\cite{tanmoy}, saturated DPPC chains may be in one of two states:
ordered and disordered. Here, we introduce a second type of lipid with
two unsaturated chains, say DOPC. These are always found in the
disordered state because the system is simulated at temperatures
higher than the low melting temperature of the DOPC. Thus, each
lattice site obtains one of five possible states depending on the
occupant: (i) a small area void ($s=0$), (ii) a disordered ($s=1$) or
(iii) an ordered ($s=2$) DPPC chain, (iv) a Chol ($s=3$), (v) a DOPC
chain ($s=4$).  In the lattice model, the detailed molecular picture
of the forces between adjacent molecules is simplified and represented
by nearest-neighbor interactions with effective parameters. The model
Hamiltonian reads

\begin{widetext}
\begin{equation}
E=-\Omega_1 k_BT\sum_{i} \delta_{s_i,1}-\sum_{i,j}\epsilon_{22}\delta_{s_i,2}\delta_{s_j,2}-\sum_{i,j} \epsilon_{23}\left[\delta_{s_i,2}\delta_{s_j,3}
  +\delta_{s_i,3}\delta_{s_j,2}\right]
-\sum_{i,j} \epsilon_{24}\left[\delta_{s_i,2}\delta_{s_j,4}
  +\delta_{s_i,4}\delta_{s_j,2}\right], \label{eq:mcenergy}
\end{equation}
\end{widetext}

where the deltas are Kronecker deltas, and the summations are carried
over all $N_s$ lattice sites in the first term and over all nearest
neighbor pair sites in the other terms.  The first term in
Eq.~(\ref{eq:mcenergy}) accounts for the fact that the disordered
state of the DPPC chains is entropically favored by a free energy,
$-\Omega_1 k_BT$ (where $k_B$ is Boltzmann constant and $T$ is the
temperature), over the ordered state. The other terms represent
effective packing attraction between an ordered DPPC chain with
another ordered DPPC chain (second term), Chol (third term), and DOPC
chains (fourth term). For the model parameters in the first three
terms in Eq.~(\ref{eq:mcenergy}), we use the same values as in
ref.~\cite{tanmoy}: $\Omega_1=3.9$, $\epsilon_{22}=1.3\epsilon$ and
$\epsilon_{23}= 0.72\epsilon$, where the energy unit $\epsilon$ is
such that the melting temperature of a pure DPPC membrane is
$T_m=314{\rm K}=0.9\epsilon/k_B$~\cite{klump,tanmoy}. The values of
the model parameters agree to order of magnitude with experimental
data for the enthalpy change at the melting transition of DPPC
vesicles.\footnote[4]{The melting transition reflects a competition
  between the disordered and ordered states of the DPPC chains. The
  former is favored because of the entropy of the chains while the
  latter packs more efficiently with the surrounding chains. One
  should not expect $\epsilon_{22}$ to be in perfect agreement with
  the melting enthalpy because this is an effective free energy
  parameter in a highly simplified model where the entropic and
  energetic influences of the implicit (missing) degrees of freedom
  are not entirely separable. Furthermore, there are entropic
  contributions that are not directly represented by the model
  parameters, e.g., the mixing entropy of the disordered lipids with
  the area voids.} The fourth term account for the affinity between
ordered DPPC and disordered DOPC chains, thereby directly influencing
their degree of mixing and controlling the line tension between the
liquid-ordered and liquid-disordered regions. We consider two
different values $\epsilon_{24}=0.4\epsilon$ and
$\epsilon_{24}=0$. The latter result in a Type II mixture that fits
nicely to the DPPC/DOPC/Chol phase diagram, while the former yields a
Type I mixture with small liquid-ordered domains that arguably may be
fitted to the very same phase diagram. Note that although we could
include interaction terms between other components (and, thus, refine
the results presented below), we prefer not to introduce such
interactions in order to keep the model minimal and highlight only the
essential biophysical driving forces.

The MC simulations employ the following move types: (i) Displacements
and rotations of randomly chosen lipids and Chol molecules.
Such moves are possible if the neighbor site is either empty or
occupied by a Chol monomer, in which case the Chol swaps places with
the displaced lipid chain, and (ii) a change in the state of a DPPC
chain (ordered to disordered and vice versa). Simulations are
conducted at temperatures of 280K or 310K, at which the area per DOPC
lipid is $\simeq 0.67~nm^2$ and $\simeq 0.73~nm^2$,
respectively~\cite{pan2008}. At this temperature range, the areas per
DPPC lipid and Chol are $\simeq 0.55~nm^2$ and $\simeq 0.28~nm^2$,
respectively~\cite{javanainen17,smond99}. We set the lattice spacing
to $l=0.56~nm$, which means that the linear size of the simulation
cell is $\simeq 70~nm$. Depending on the composition of the simulated
system, we set the number of voids such that the average density of
the system matches the weighted average density of the constituting
molecules. Systems are equilibrated for $10^{10}$ MC trail moves (95\%
of which are displacements and rotations and 5\% are state exchange)
until the memory of the initial configuration (either a random or a
highly ordered distribution of the molecules on the lattice) is lost
and the energy of the system relaxes to its characteristic value. We
then continue to sample the system for $10^{11}$ MC trial moves during
which we measure quantities of interest.

\begin{figure*}[t]
\centering
\includegraphics[width=16cm]{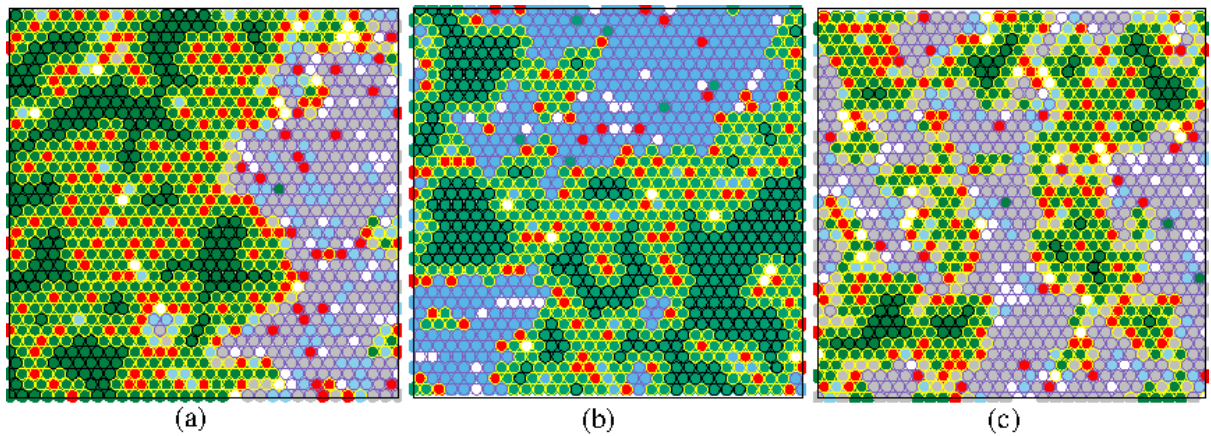}
\caption{Magnified views of small portions of (a) Type II ternary
  mixture, (b) binary mixture (adapted from ref.~\cite{tanmoy}), and
  (c) Type I ternary mixture. Ordered DPPC, disordered DPPC, DOPC,
  Chol, and empty sites are colored in green, light blue, grey, red,
  and white, respectively. Purple, yellow and black borders are
  assigned to the sites from liquid-disordered, liquid-ordered, and
  gel regions, respectively.}
\label{fig2}
\end{figure*}

In the next section we present snapshots (see Figs.~\ref{fig1},
\ref{fig3}, \ref{fig4}, and \ref{fig6}) where the sites of the lattice
are drawn with distinct colors, denoting their belonging to
liquid-disordered, liquid-ordered, and gel regions of the system. The
partition of the sites into different categories is based in the
following simple algorithm: The sites may be in one of five states:
void, disordered DPPC chain, ordered DPPC chain, Chol, and DOPC
chain. Each site is assigned with a score representing the degree of
order of the associated state. Explicitly, liquid-ordered domains are
rich in ordered DPPC and Chol, which receive positive scores of
$Sc_i=2$ and $Sc_i=1$, respectively. Conversely, the liquid-disordered
regions are populated with DOPC and disordered DPPC chains, which get
negative scores of $Sc_i=-1$, and $Sc_i=-0.5$, respectively. Voids
have a zero score. The {\em grade}\/ of a site is given by
\begin{equation}
  G_i=Sc_i+\sum_{j=1}^6 Sc_j
  \label{eq:grade}
\end{equation}
where the sum runs over the six nearest neighbors of the site. Sites
with non-negative (negative) grades, $G_i\ge 0$ ($G_i<0$), are
associated with the liquid-ordered (liquid-disordered) regions. The
maximum grade of $G_i=14$ is obtained for ordered DPPC chains
surrounded by 6 other ordered DPPC chains. These sites are marked as
gel. After dividing the sites to the liquid-disordered,
liquid-ordered, and gel regions, we scan the lattice one extra time
and switch a liquid-disordered site to liquid-ordered if: (i) it is
occupied by an ordered DPPC chain, and (ii) it has at least one
neighbor belonging to a liquid-ordered or gel regions. The final
iteration is needed in order to include ordered chains located at the
periphery of the liquid-ordered domains that are surrounded by many
disordered chains and incorrectly classified as part of the
liquid-disordered regions.

For a given composition of components, we have generated many
configurations starting at different initial distributions. We
simulated the mixtures until they relax into states representing
equilibrium and verified that different initial conditions evolve into
similar phases: For instance, the snapshot in Fig.~\ref{fig1}(b) shows
a macroscopically large liquid-ordered domain in a liquid-disordered
membrane (where some small liquid-ordered domains are also
floating). This snapshot can be unambiguously intinterpeted as an
indication that the equilibrium state of the mixture is a macroscopic
$L_d+L_o$ phase coexistence. All other independent MC runs with the
same compositions also evolve into similar structures corresponding to
macroscopically phase separation between the two liquid phases. Our
goal in this study is not to plot the exact locations of the
boundaries of the phase diagram. These are adopted from
experiments~\cite{veatch07}, and we investigate whether the outcomes
of the simulations are consistent with the experimental phase
diagram. Indeed, a visual inspection of the equilibrated systems shows
good consistency with the experimentally determined phase diagram.
Our lattice simulations results also show reasonable agreement with
atomistic simulations data for the compositions of the
liquid-disordered and liquid-ordered regions (see Table~\ref{table:1}
below).

\section*{Results} 

\subsection*{Type II DPPC/DOPC/Chol mixtures}

Setting $\epsilon_{24}=0$ in Eq.~(\ref{eq:mcenergy}) for the DOPC
interactions while keeping the values of the other model parameters as
in our previous work on DPPC/Chol binary mixture~\cite{tanmoy}, yields
a ternary DPPC/DOPC/Chol Type II mixture model that reproduces the
anticipated phase diagram.  Our simulations at $T=280$K are summarized
in Fig.~\ref{fig1}. Fig.~\ref{fig1}(a)-(h) show representative
equilibrated snapshots from simulations with different compositions,
the location of which in the ternary phase diagram are marked in red
in Fig.~\ref{fig1}(i). The purple points in Fig.~\ref{fig1}(i)
indicate additional simulated compositions whose snapshots are not
shown.  In the snapshots, we color sites of the liquid-disordered,
liquid-ordered, and gel regions with purple, yellow, and black,
respectively. The blue lines in the phase diagram mark the
experimentally-determined phase boundaries of DPPC/DOPC/Chol mixtures,
as adapted from ref.~\cite{veatch07}. We note that our aim herein is
not to determine the phase lines but demonstrate that the model yield
results that fit them. Indeed, we see in Fig.~\ref{fig1} that the
snapshot configurations fit very well into the different regions in
the phase diagram. DPPC/DOPC/Chol is a Type II mixture and, indeed, we
observe in the two- and three-phase regions that the different phases
are macroscopically segregated. At the microscopic scale, we notice
that the different phases are locally inhomogeneous. This is best seen
in fig~\ref{fig1}(h) displaying a single inhomogeneous phase with
liquid-ordered domains floating in a liquid-disordered sea. This
near-critical arrangement, clearly differs from the $L_d+L_o$
coexistence appearing in snapshot (b).

Another example of a local inhomgeneity is seen in snapshot (g) of the
$L_o$ phase, where small gel clusters appear in the liquid-ordered
region. Conversely, in the gel phase in snapshot (e), we see local
liquid-ordered impurities. These inhomogeneities, which are also seen
in snapshots (b) and (d) (where the $L_0$ and gel phases coexist,
respectively, with the $L_d$ phase) represent local variations in the
composition of the ordered regions. Recall that the sites marked by
black color in the snapshots are the sites occupied by ordered DPPC
chains with six ordered DPPC nearest neighbors. Since the gel phase in
(e) also contains small amounts of Chol, not all of its sites, only
most of them, can satisfy this criterion.  Another notable observation
in snapshot (e) of the gel phase is the segregation of a mesoscopic
liquid-disordered domain containing mostly DOPC lipids. The presence
of islands of disordered lipids within the ordered phases has also
been observed experimentally~\cite{veatch05}. The fluorescence
microscopy micrographs appear as uniform $L_o$ or $S_o$ phases only
when no DOPC is present. This experimental observation indicates that
the unsaturated lipid has extremely low miscibility in the ordered
phases. Despite the smallness of the liquid-ordered region, it must be
admitted that it is more natural to interpret the appearance of a DOPC
cluster in snapshot (e) as coexistence between gel and $L_d$ phases,
i.e., similarly to snapshot (d) but with different fractions of DOPC
and DPPC. We note here that a single small liquid-disordered island
was observed in simulations of mixtures with even lower amounts of
DOPC (not shown), which raises the concern that our model parameters
may not be accurately tuned and therefore result in excessively low
miscibility of the DOPC lipids in the gel phase. In contrast to (e),
in the $L_o$ phase in snapshot (g), the segregation tendency of DOPC
appears to be weaker and they form several smaller domains whose
typical size falls below microscopic resolution. This looks like a
mirror image of snapshot (h), and so the model adequately describes
the continuous crossover between the $L_d$ and $L_0$ phases at
moderate Chol concentrations.

Snapshot (f) exhibits intermediate properties of both the gel (e) and
the $L_o$ (g) phases, in terms of the relative fractions of gel and
non-gel regions and the degree of segregation of the disordered
part. One can therefore classify snapshot (f) as $L_o+S_o$ coexistence
phase, although the separation is local rather than macroscopic
(notice that the distribution of gel clusters inside the ordered
region is locally inhomogeneous), and there seem to be a continuous
crossover rather than a discontinuous transition between the $L_o$ to
gel phases.  Likewise, snapshot (c) is recognized as a $L_d+L_o+S_o$
because of the intermediate nature between snapshots (b) and (d). We
emphasize that the ability of our simulation snapshots to separate the
disordered $L_d$ phase from the ordered ones, while struggling to
distinguish between the more similar ordered $L_o$ and $S_o$ phases is
fully consistent with fluorescence microscopy results where the same
problem has been encountered~\cite{veatch05}. Indeed, it has been
noted that fluorescence microscopy alone is not well suited for
studying the onset of the gel phase, and other experimental methods
that are sensitive to smaller-scale lipid organization, e.g., NMR
spectra, should be better suited for characterizing this putative
three phase region. Our simulation results provide a picture of this
local organization, suggesting that in the three phase region the
$L_d$ phase is macroscopically separated, while the $L_o$ and $S_o$
phases are segregated only at the local scales. In this context, it is
important to remind that small gel clusters residing within the $L_o$
phase have been also detected inside liquid-ordered domains in binary
DPPC/Chol mixtures, suggesting that the thermodynamic origin of this
observation is similar in both systems. This can be understood
considering that the first three terms in Eq.~(\ref{eq:mcenergy})
constitute the binary mixture Hamiltonian in ref.~\cite{tanmoy}, and
that we set to zero the coefficient in the fourth term accounting for
the DOPC interactions. Thus, the picture inside the $L_o$ phase remain
quite similar, as demonstrated in Fig.~\ref{fig2} showing smaller
portions of different simulated systems. Fig.~\ref{fig2}(a) is a
magnification of the region marked by a red square in snapshot (b) of
Fig.~\ref{fig1}. Depending on the state of the lattice site, it is
marked by the following color: white ($s=0$, void), light blue ($s=1$,
disordered DPPC chain), green ($s=2$, ordered DPPC chain), red ($s=3$,
Chol), and grey ($s=4$, DOPC chain). Additionally, the boarders of the
sites are marked with the same color coding as in the snapshots in
Fig.~\ref{fig1} in order to indicate in which region
(liquid-disordered - purple, liquid-ordered - yellow, and gel - black)
they reside. Fig.~\ref{fig2}(b) shows a magnified view from our
previous simulations of DPPC/Chol mixtures~\cite{tanmoy}, featuring
similar local distributions of gel nano domains inside the
liquid-ordered region. Fig.~\ref{fig2}(c) is taken from our
simulations of Type I ternary mixtures, to be presented in the
following section [see red box in Fig.~\ref{fig4}(b) below]. In Type I
mixtures, the domains are smaller but, nevertheless, exhibit, similar
local hetrogeneities. The thermodynamic origin of these arrangements
is the strong packing attraction between the ordered DPPC which drives
their clustering into small gel domains. Chol, which prefers the
vicinity of ordered DPPC chains (and, therefore, induces ordering of
DPPC) is expelled from these clusters because its attraction to them
is weaker then their mutual attraction. It accumulates in the
liquid-ordered region, where it acts as a sort of a lineactant between
the gel and the liquid-disordered region of the system.

\begin{figure}[h]
\centering
\includegraphics[width=6cm]{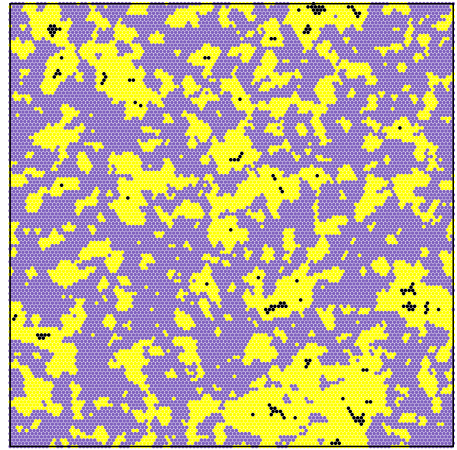}
\caption{A snapshot from the Type II mixture simulations at
  DPPC/DOPC/Chol mole fraction composition of 0.23:0.38:0.39 and
  $T=280$K. Color coding is as in the snapshots in Fig.~\ref{fig1}.}
\label{fig3}
\end{figure}

The location of a second order critical point on the binodal curve
separating the one phase and the two-phase $L_d+L_o$ regions is
roughly near the intersection of binodal curve with the line
corresponding to DOPC mole fraction, $\phi_{\rm DOPC}=0.4$ [along
  which points (h) and (b)
  reside]~\cite{veatch07,veatch03,davis,uppamoochikkal}.  The
incomplete segregation of the system in (h) into small liquid-ordered
domains and a liquid-disordered matrix is viewed as a precursor to the
phase transition.  According to this picture, the density fluctuations
in the one phase region are expected to intensify in the proximity of
the critical point, leading to a broader distribution of domain sizes
and the emergence of a percolation liquid-ordered cluster. Evidences
supporting this scenario may be found in Fig.~\ref{fig3} showing a
typical configuration DPPC/DOPC/Chol at mole fraction composition of
0.23:0.38:0.39. The distribution of liquid-ordered domain sizes is
indeed broader here compared to Fig.~\ref{fig1}(h) which, presumably,
is located further from the critical point. Moreover, one can notice
that the liquid-ordered domains in Fig.~\ref{fig3} almost percolate
across the lattice.

\begin{figure*}
\centering
\includegraphics[width=16cm]{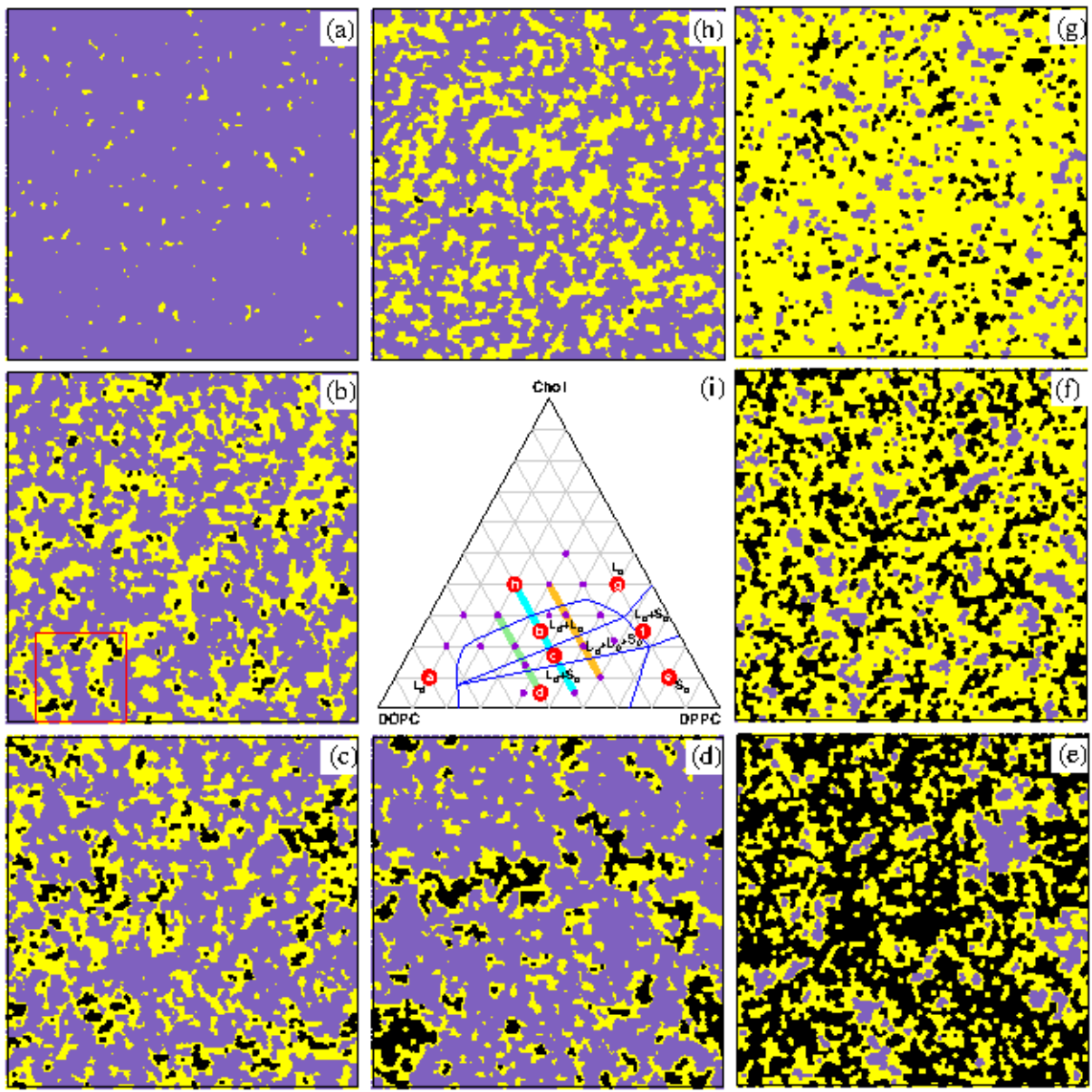}
\caption{Same as Fig.~\ref{fig1} but for $\epsilon_{24}=0.4\epsilon$.}
\label{fig4}
\end{figure*}

To conclude this section, the simulation snapshots agree nicely with
the experimental phase diagram in regard to one phase ($L_d$, $L_o$)
and two-phase $L_d+L_o$ regions, including the estimated location of
the second order critical point. As for the $S_o$ gel phase, the
miscibility of DOPC in this phase may be underestimated by the model
which causes them, in the simulations, to phase separate and form
small $L_d$ domains that might be below macroscopic resolution. An
interesting conclusion drawn from our simulations is the fact that the
two ordered phases ($L_o$, $S_o$) exhibit local but not macroscopic
separation. This observation is also consistent with experimental
data, where $L_o$+$S_o$ coexistence has been inferred from
spectroscopic methods that are sensitive to local details. In the
following section dealing with Type I mixtures, we expand on the
connection between scattering data and the detection of
locally-separated phases in ternary mixtures.

\subsection*{Type I version of DPPC/DOPC/Chol mixtures}

In the previous section we saw that in Type II mixtures, like
DPPC/DOPC/Chol, small liquid-ordered domains are expected in the one
phase region, i.e., above the binodal line but not too far from the
critical point. But what about Type I mixtures that do not exhibit a
true thermodynamic coexistence? In this section we demonstrate that
such mixtures may still feature small liquid-ordered domains that
float in the sea of a liquid-disordered membrane. Explicitly, we
consider the same system as in the previous section and only change
the value of the model parameter $\epsilon_{24}$ from 0 to 0.4,
thereby allowing better mixing of the ordered DPPC and disordered DOPC
chains. As we will now see, for this value of $\epsilon_{24}$, the
mixture may arguably be regarded as a Type I version of the Type II
DPPC/DOPC/Chol mixture studied in the previous section: It lacks
thermodynamic phase transition on the one hand, but fits the
DPPC/DOPC/Chol phase diagram {\em if local measures are used to
  distinguish between the one phase and two phase regions}.

Fig.~\ref{fig4} summarizes our simulation results at $T=280$K for
$\epsilon_{24}=0.4\epsilon$. The color coding used here is similar to
the one in Fig.~\ref{fig1}, and the phase diagram in (i) is identical
to Fig.~\ref{fig1}(i), i.e., drawn with the blue phase boundaries
taken from~\cite{veatch07}.  One clearly sees that the change in the
value of $\epsilon_{24}$ eliminates any signature of macroscopic phase
separation, leaving only evidences of local phase separation, e.g., in
snapshots (h), (b), (c), and (d). These snapshots indicate that the
mixture is now of Type I.  The question to be addressed now is whether
the simulation snapshots in Figs.~\ref{fig4}(a-h) still fit to the
phase diagram in Fig.~\ref{fig4}(i), which is the phase diagram of the
Type II DPPC/DOPC/Chol ternary mixture. We first note that snapshots
(a), (g), and (e) are consistent with the expectations to observe
single $L_d$, $L_o$, and $S_o$ phases at the respective compositions.
Identifying a two phase region when the system displays no macroscopic
phase separation is, of course, harder. Snapshot (f), presumably
showing coexisting $L_o$ and $S_o$ regions, exemplifies why the
interpretation of the results is ambiguous. Recall that the gel
regions are hexagonally packed arrangements of ordered DPPC
chains. Therefore, we associate the gel phase in the snapshots with
the lattice sites containing DPPC ordered chains with six nearest
neighbors of the same kind. Both the liquid-ordered and gel phases
appearing in (g) and (e), respectively, are highly populated with
ordered DPPC chains. They differ in the fractions of chains obeying
the six nearest neighbor criterion, which is high in (e) and low in
(g). Looking at snapshot (f), one can interpret the observed
distribution as a coexistence of regions of $L_o$ and $S_o$ phases,
which are separated locally rather than macroscopically because of the
smallness of the line tension between them. It is, however, impossible
to rule out that the sequence of snapshots (e)-(f)-(g) simply shows a
gradual decrease in the fraction of sites classified as gel. In this
scenario, snapshot (f) is interpreted as a single phase which is
locally inhomogeneous with regions that look more gel-like and regions
resembling the liquid-ordered phase.

The main question pertains to the coexistence region of two liquid
phases, $L_d+L_o$. As noted above, the justification for drawing the
binodal curve in the phase diagram (i) is based on an ambiguous
interpretation of scattering data. The detection of small domains
below the resolution of optical microscopy is based on signals of
local order in, e.g., FRET, NMR and electron spin resonance (ESR)
spectroscopy~\cite{heberle10,yasuda15}, but the results of such
measurement do not resolve the question whether the ordered domains
represent genuine phases or not. The problem is nicely exemplified by
snapshots (h) and (b) which display visually similar distributions of
liquid-disordered and liquid-ordered domains. These snapshots resemble
the near-critical Type II mixture appearing in Fig.~\ref{fig3}. The
fact that we do not observe macroscopic phase coexistence in the
present case does not necessarily rule out the possibility that
snapshot (b) displays two-phase coexistence. This is because of the
usual problem to distinguish between a single inhomogeneous phase and
two-phase coexistence in the critical region where the mixture
exhibits large thermal density fluctuations. Keep in mind that the
snapshots in Fig.~\ref{fig4} show the classification of the lattice
sites to different states (liquid-disordered/liquid-ordered/gel), but
provide no real information about the local density
fluctuations. Therefore, the boundaries between the ordered and
disordered regions are probably fuzzier than appearing in the
figures. With that said, we note that our state-classification
algorithm works well and yields very good consistency with data from
ref.~\cite{tieleman20} reporting on atomistic simulations of a Type I
version of DPPC/DOPC/Chol mixtures. Table~\ref{table:1} lists the
average compositions of the liquid-disordered and liquid-ordered
domains, measured for mixtures with overall DPPC/DOPC/Chol mole
fraction of 0.35:0.35:0.30 (located in the $L_d+L_o$ region of the
phase space). The agreement between the lattice and atomistic
simulations is more than fair, even though that the domain
classification algorithms used here and in ref.~\cite{tieleman20} are
different. A visual inspection of different snapshots indicates that
the domain sizes in both works are also similar. We note here that it
is not clear why the atomistic simulations fail to exhibit the
experimentally-observed macroscopic phase separation of DPPC/DOPC/Chol
mixtures. It may be due to a problem with the force fields or an
artifact of the small system size. We nevertheless use them as a
benchmark for our simulations of the Type I version of DPPC/DOPC/Chol
mixtures.

\begin{table}
\small
\caption{Comparisons of lattice (this work) and atomistic
  (ref.~\cite{tieleman20}) simulations data for the average
  compositions of the liquid-disordered and liquid-ordered domains for mixtures with
  DPPC/DOPC/Chol mole fraction 0.35:0.35:0.30.}
\begin{tabular}{|c|c|c|c|}
\hline
\multicolumn{4}{|c|}{DPPC/DOPC/Chol composition} \\
\hline
Temperature (K) & model & liquid-ordered & liquid-disordered\\
\hline
\multirow{2}{*}{280} & Lattice & 0.46:0.16:0.38 & 0.15:0.67:0.18\\
\cline{2-4}
 & Atomistic & 0.53:0.13:0.34 & 0.16:0.60:0.24\\
\hline
\multirow{2}{*}{310} & Lattice & 0.44:0.15:0.41 & 0.26:0.55:0.19\\
\cline{2-4}
 & Atomistic & 0.48:0.14:0.38 & 0.20:0.60:0.20\\
\hline
\end{tabular}
\label{table:1}
\end{table}

\begin{figure*}[t]
\centering
\includegraphics[width=16cm]{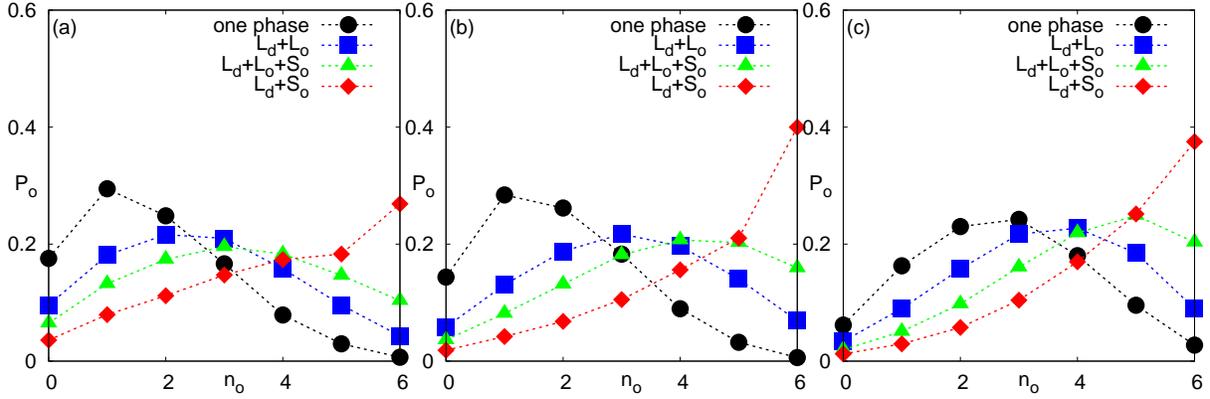}
\caption{The probability $P_o$ of finding an ordered DPPC chain in the
  liquid-ordered and gel regions with $n_o$ ordered DPPC nearest
  neighbors. Data in (a), (b), and (c) corresponding to simulations
  with $\phi_{\rm DOPC}=$ 0.5, 0.4, and 0.3, respectively. For each
  $\phi_{\rm DOPC}$, the statistics is calculated at four different
  compositions of systems belonging to the one phase (data plotted
  with black circles), two phase $L_d+L_o$ (blue squares), three phase
  $L_d+L_o+S_o$ (green triangles), and two phase $L_d+S_o$ (red
  diamond) parts of the phase space.
  Lines are guides to the eyes.}
\label{fig5}
\end{figure*}

The problem lies not in the accuracy of the algorithm for classifying
sites to liquid-disordered, liquid-ordered, and gel regions, but in
its resolution.
The algorithm, which is based on the inspection of the local vicinity of each
lattice site, is too coarse 
to resolve the differences between configurations (h) and (b), and one
must inspect more closely (i.e., define additional measures to
quantify) the distribution of lipids at the nanometric small scales.
In our model, we can gain such insight from the statistics of the
ordered DPPC chains that are classified by the algorithm as either in
the liquid-ordered or gel regions. (Ordered DPPC chains barely exist
in the liquid-disordered region.)  Explicitly, we check all these
ordered DPPC chains and compute of the probability $P_o$ of these
chains to have $n_o=0,1,2,3,4,5,6$ other ordered DPPC neighbors. Our
analysis is summarized in Fig.~\ref{fig5} showing the statistics at
four different points along the green, cyan, and orange lines
appearing in Fig.~\ref{fig4}(i). These lines, corresponding
respectively to $\phi_{\rm DOPC}=0.5$ [statistics plotted in
  Fig.~\ref{fig5}(a)], $\phi_{\rm DOPC}=0.4 $ [Fig.~\ref{fig5}(b)],
and $\phi_{\rm DOPC}=0.3$ [Fig.~\ref{fig5}(c)], cross four different
parts of the phase space: the one phase region, two phase $L_d+L_o$,
three phase $L_d+L_o+S_o$, and two phase $L_d+S_o$.  The data in
Fig.~\ref{fig5} suggests that the most distinguishing feature between
configurations in the one phase and the $L_d+L_o$ two phase regions is
the probability of having six ordered neighbors, which is quite
negligible in the former and somewhat more substantial in the latter.
This suggests that the presence of small gel clusters in the ordered
domains is a defining property of the $L_o$ phase. In the $L_d+S_o$
region, $P_o$ is a monotonically increasing function of $n_o$. This is
a characteristic feature of the gel phase.  In the three phase region,
the probability function exhibits features of both the $L_d+L_o$ and
$L_d+S_o$ probabilities. As in the former, $P_o$ is {\em not}\/
monotonically increasing; but similarly to the latter, there is a
significant probability to detect an ordered chain with $n_o=6$
neighbors, indicating that the gel phase is also present.

\begin{figure}
\centering
\includegraphics[width=7.5cm]{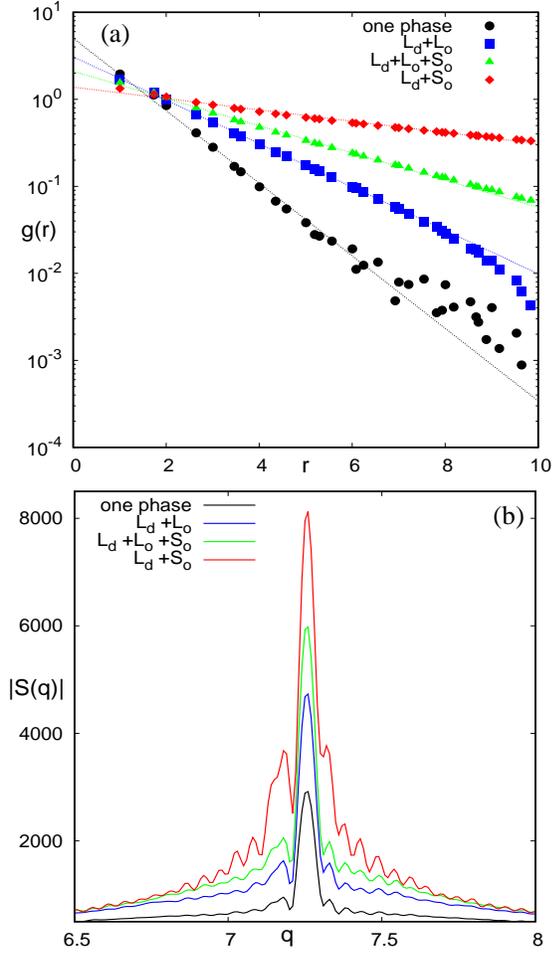}
\caption{(a) The pair correlation function defined in
  Eq.~(\ref{eq:pair}) as a function of the pair distance $r$ (measured
  in units of the lattice spacing $l$), and (b) the associated model
  structure factor Eq.~(\ref{eq:factor}), as a function of the wave
  vector $q=|\vec{q}|$. The data corresponds to simulations of four
  mixtures with $\phi_{\rm DOPC}=0.4$ situated in the one phase (data
  plotted with black circles), two phase $L_d+L_o$ (blue squares),
  three phase $L_d+L_o+S_o$ (green triangles), and two phase $L_d+S_o$
  (red diamond) regions of the phase diagram. Lines are guides to the
  eyes.}
\label{fig6}
\end{figure}

To make an even closer connection with experimental data and provide
more insight into the differences between regions of the phase
diagram, we computed the spatial pair correlation function of local
orderliness $g(r)$ and the associate Fourier transformation
$|S(q)|$. More precisely, the spatial correlation function is
calculated from the grade function $G_i$ used for the color coding of
the snapshots in Figs.~\ref{fig1} and~\ref{fig4}. Recall that the
grade of a lattice point $G_i=G(x_i)$ is defined by
Eq.~(\ref{eq:grade}) and assumes positive (negative) values for
lattice points in the ordered (disordered) regions. To focus on the
size of the ordered domains, we reset the grades of the disordered
lattice points to zero, and calculate
  \begin{equation}
  g(r)\equiv\frac{\left\langle G(x_i)G(x_i+r)\right\rangle}{\left\langle
    G(x_i)\right\rangle^2}-1,
  \label{eq:pair}
  \end{equation}
where $r$ is the pair distance between lattice points, and the
averages are taken over all the lattice points $x_i$ and over 100
independent snapshots of the system. The analysis is conducted at the
same four compositions studied in Fig.~\ref{fig5}(b) above, and
summarized in Fig.~\ref{fig6}(a). For all four compositions, which are
located along the cyan line in different regions of the phase diagram
Fig.~\ref{fig4}(i), the function $g(r)$ is well fitted by a single
exponential decaying function. The correlation length, which is
inversely proportional to the slope of the lines in
Fig.~\ref{fig6}(a), serves as a measure for the characteristic linear
dimension of the ordered domains. As expected from a visual inspection
of snapshots, our results show that the size of the ordered domains
grow as the composition of the mixture is varied from the one phase
region of the phase diagram to the $L_d+S_o$ region.

In reality, the pair correlation function of solids and dense liquids
show oscillations at small distances because of the packing of the
molecules. These oscillation are obviously absent in
Fig.~\ref{fig6}(a) because our data is taken from lattice
configurations that provide no information at distances not
corresponding to one of the lattice vectors. Therefore,
Fig.~\ref{fig6}(a) shows the envelope of the pair correlation
function, from which we can extract information on the size of the
domains. Insight into the internal domain structure and correlations
can be gained from the model structure factor, $|S(q)|$, defined by
calculating the Fourier sum of the grade function $S(\vec{q})=\sum_i
G(x_i)e^{i\vec{q}\cdot\vec{x_i}}$, and then taking the angular average
over all the orientations of the vector of size
$\vec{q}=(q\cos(\theta),q\sin(\theta))$\footnote[2]{Because our
  simulations are conducted on a two-dimensional lattice we only take
  the average over the azimuthal angle. In X-ray scattering
  experiments of three-dimensional solutions, the average is over both
  the polar and azimuthal angles.}:
  \begin{equation}
    |S(q)|=\int_0^{2\pi}|S(\vec{q})|d\theta.
    \label{eq:factor}
  \end{equation}
 The model structure factor is plotted in Fig.~\ref{fig6}(b) as a
 function of the wave-vector $q$. For all four mixture compositions
 studied here, the central peak is located at the reciprocal lattice
 wave vector $q^*=4\pi/\sqrt{3}$ (in units of inverse lattice spacing
 $l^{-1}$). The magnitude of the peak increases when the composition
 of the mixture is varied from the one phase region to the $L_d+S_o$
 region, reflecting the growth of the characteristic domain size along
 which the spatial correlation is maintained as seen in
 Fig.~\ref{fig6}(a). Another trend observed in Fig.~\ref{fig6}(b) is
 the increase in the amplitude and number of the satellite peaks
 around $q^*$. The satellite peaks serve as indicators for modulations
 in the orderliness of the domains. We can identify in
 Fig.~\ref{fig6}(b) the differences between the phase space regions:
 In the one phase region we find a single weak satellite peak
 originating from the small domains floating in the mostly disordered
 mixture. In the $L_d+L_o$ region, the amplitude of this peak grows
 and a few more weak peaks appear. This is the contribution of the
 liquid ordered domains, especially the small gel-like regions inside
 them. When the gel clusters occupy a more substantial portion of the
 system, the number and amplitude of the satellite peaks increase
 significantly. We note here that we have verified (by visual
 inspection) that the appearance of many satellite peaks in the
 $L_d+S_o$ region does not arise because the system is frozen.

\subsection*{Type II mixture at physiological temperature}

\begin{figure*}
\centering
\includegraphics[width=16cm]{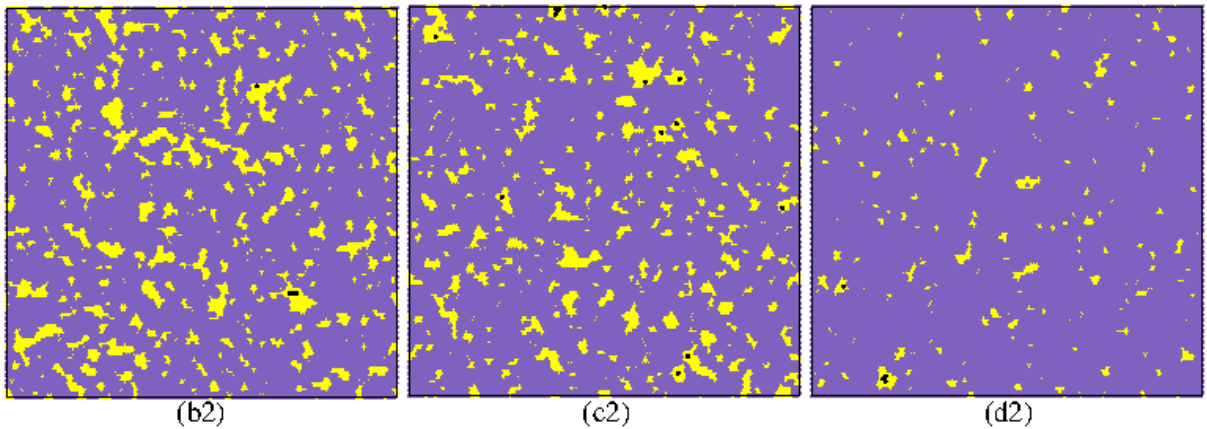}
\caption{Snapshots of equilibrated mixtures at temperature $T=310$ K
  with $\epsilon_{24}=0$.  The compositions in (b2), (c2), and (d2)
  are the same as in (b), (c), and (d) in Fig.~\ref{fig1},
  respectively. Color coding is similar to Fig.~\ref{fig1}.}
\label{fig7}
\end{figure*}

At $T=310$K, most of the DPPC chains are found in the disordered
state, which is understandable considering the proximity to their
melting temperature $T_m=314$K. At this temperature, a single $L_d$
phase is expected throughout the entire phase diagram except, perhaps,
for mixtures with a very high concentrations of
DPPC~\cite{veatch07}. Nevertheless, this phase may still contain small
liquid-ordered domains, as demonstrated in Fig.~\ref{fig7}, showing
equilibrated configurations at compositions similar to points (b),
(c), and (d) in Fig.~\ref{fig1} [denoted in Fig.~\ref{fig7} by (b2),
  (c2), (d2), respectively]. This is an important observation if one
considers the ternary mixture model as a framework for understanding
the formation of raft domains in biological membranes. As the
snapshots in Fig.~\ref{fig7} demonstrate, our minimal model is capable
of producing liquid-ordered domains of sizes $\lesssim 10$ nm even at
the physiological temperature. These are local inhomogeneities with a
characteristic correlational length which is half to one order of
magnitude smaller than rafts in biological membrane, which is a well
known limitation of experimental ternary
mixtures~\cite{schmid17review}.  This is why additional factors and
mechanisms, e.g., curvature composition
coupling~\cite{leibler,schick1,schick2,sadeghi}, as well as the
presence of specific proteins and the cell cytoskeleton~\cite{kusumi,
  fischer}, are believed to play a role in stabilizing larger
domains. However, our results suggest that the same effect of
increasing the domain sizes may be also accomplished by tuning
differently the model parameters. After all, biological membranes are
{\em not}\/ ternary mixtures. They are composed of many lipid species
with different melting temperatures and packing interactions, and our
study
suggests that even small changes in these parameters may affect
dramatically the size distribution of the liquid-ordered domains.

\section*{Discussion}

We presented a simple lattice model for ternary mixtures of saturated
and unsaturated lipids with Chol, involving only a minimal number of
effective nearest-neighbor interactions between the constituting
components. The model succeeds in explaining experimental and
atomistic simulation observations across multiple scale, ranging from
the local distributions of lipids to the macroscopic phase
diagram. The model Hamiltonian is constructed by adding a single
interaction term, between unsaturated (DOPC) and saturated lipids, to
a previously presented model of a binary mixture of saturated lipids
(DPPC) and Chol. The minimal nature of the model allows us to identify
the influences of the different effective packing interactions and the
underling demixing mechanisms. The formation of the $L_o$ phase is
obviously related to the affinity of Chol to the saturated
lipids. This phase is inhomogeneous and contains gel-like nano-clusters
whose origin is the particularly strong attraction of the saturated
lipids to each other. The strength of the interaction between the
unsaturated and saturated lipids, $\epsilon_{24}$, controls the nature
of phase transitions of the ternary mixture and the sizes of the
liquid-ordered domains. For $\epsilon_{24}=0$, the simulated mixture
exhibits a macroscopic phase separation and fits nicely to the phase
diagram of a Type II DPPC/DOPC/Chol mixture. In this case,
liquid-ordered domains are observed in the one-phase region near the
critical point of the $L_d+L_o$ coexistence binodal. For
$\epsilon_{24}=0.4$, we obtain the Type I version of a DPPC/DOPC/Chol
mixture with nanoscopic domains. This system can be also fitted to the
DPPC/DOPC/Chol phase diagram provided that local measures, such as the
statistics of the ordered DPPC chains, are considered. In scattering
experiments, nanometric features of this kind leave signatures in the
scattering data, thus making the interpretation of configurations like
in Fig.~\ref{fig4}(b) ambiguous - either as coexistence of two liquid
phases separated by a very small line tension or as a single phase
with local inhomogeneities. Our aim here was not to decide which
interpretation is correct, but to demonstrate that even if the phase
transitions of the Type II DPPC/DOPC/Chol mixture are lost in the Type
I version, local traces of phase coexistence still remain.

The motivation for studying lipid/Chol mixtures with a few components
largely stems from the belief that they provide insight into the
thermodynamic properties of cellular membranes. Because biological
membranes are much complex physical systems, one should be cautious
when coming to draw conclusions on their behavior from simpler model
systems. On the one hand, the fact that biological membranes are
mixtures of many different types of lipids suggests that it may be
possible to construct a suitable model with many interaction terms
that not only shows formation of liquid-ordered domains, but also
reproduces their size distributions (in contrast to ternary mixtures
where the domains are usually smaller than biological rafts). Several
recent theoretical studies of multi-component mixtures have indeed
demonstrated that strong affinity of certain components to each other
may lead to demixing phase transitions that are more robust with
respect to variations in temperature and the intermolecular
interactions than mixtures with a few components
~\cite{girard21,carugno22,jacobs17}. On the other hand, it would be
probably wrong to ignore other mechanisms beyond short-range
packing. This is particularly true for the inner cytoplasmic layer of
cellular membranes that contains a low concentration of saturated
lipids~\cite{asymmetric}.  We can expect rafts to be influenced by the
proximity to the cytoskeleton, the presence of proteins with
affinities to certain lipids, and the curvature elasticity of the
membrane (that depends on the local composition and induces coupling
between the layers via the spontaneous curvature). Assessing the
importance of these different mechanisms and understanding the
interplay between them is going to remain an open question in the
foreseeable future.





\section*{Acknowledgments}
  This work was supported by the Israel Science Foundation (ISF), grant
  No.~991/17. TS thanks the Planning and Budgeting Committee of the
  Council for Higher Education (Israel) for supporting his
  post-doctoral fellowship.

\renewcommand{\refname}{{Bibliography}}

\end{document}